\documentclass[aps,showkeys]{revtex4}

\usepackage{epsfig}
\usepackage{eepic}
\usepackage{latexsym}
\usepackage{rotating}
\usepackage{lscape}
\usepackage{graphics}
\usepackage{dcolumn}
\usepackage{bm}
\newcommand{\beq}{\begin{equation}}
\newcommand{\eeq}{\end{equation}}
\newcommand{\bd}{\begin{displaymath}}
\newcommand{\ed}{\end{displaymath}}

\newcommand{\bei}{\begin{itemize}}
\newcommand{\eei}{\end{itemize}}

\newcommand{\bee}{\begin{enumerate}}
\newcommand{\eee}{\end{enumerate}}

\begin{document}

\noindent
\title{Testing Monte Carlo absolute dosimetry formalisms for a small field `D'-shaped collimator used in retinoblastoma external beam radiotherapy}

\author{P.~A.~Mayorga$^{1}$, L.~Brualla$^2$, A.~Fl\"uhs$^2$, W.~Sauerwein$^2$ and A.~M.~Lallena$^3$\\
{\small {\it
$^1$Departamento de F\'{\i}sica, Universidad Nacional de Colombia, Cr 45 \# 26--85, Bogot\'a D.C., Colombia.\\
$^2$NCTeam, Strahlenklinik, Universit\"atsklinikum Essen, Hufelandstra\ss e 55, D-45122 Essen, Germany.\\
$^3$Departamento de F\'{\i}sica At\'omica, Molecular y Nuclear, Universidad de Granada, E-18071 Granada, Spain.
}}}

\date{\today}

\bigskip

\begin{abstract}
\noindent
{\it Purpose:} To investigate the validity of two Monte Carlo simulation absolute dosimetry approaches in the case of a small field dedicated `D'-shaped collimator used for the retinoblastoma treatment with external photon beam radiotherapy.\\
{\it Methods:} The Monte Carlo code {\sc penelope} is used to simulate the linac, the dedicated collimator and a water phantom. The absolute doses (in Gy per monitor unit) for the field sizes considered are obtained within the approach of Popescu {\it et al.} in which the tallied backscattered dose in the monitor chamber is accounted for. 
The results are compared to experimental data, to those found with a simpler Monte Carlo approximation for the calculation of absolute doses and to those provided by the analytical anisotropic algorithm. Our analysis allows for the study of the simulation tracking parameters. Two sets of parameters have been considered for the simulation of the particle transport in the linac target.\\ 
{\it Results:} The change in the tracking parameters produced non-negligible differences, of about 10\% or larger, in the doses estimated in reference conditions. The Monte Carlo results for the absolute doses  differ from the experimental ones by 2.6\% and 1.7\% for the two parameter sets for the collimator geometries analyzed. For the studied fields, the simpler approach produces absolute doses that are statistically compatible with those obtained with the approach of Popescu {\it et al.} The analytical anisotropic algorithm underestimates the experimental absolute doses with discrepancies larger than those found for Monte Carlo results.\\ 
{\it Conclusions:} The approach studied can be considered for absolute dosimetry in the case of 
small, `D'-shaped and off-axis radiation fields. However, a detailed description of the radiation transport in the linac target is mandatory for an accurate absolute dosimetry.
\end{abstract}


\maketitle

\section{Introduction}

Nowadays, the Monte Carlo (MC) simulation of radiation transport is considered to provide the most accurate determination of the energy deposited by ionizing radiation in a medium. The MC method permits an accurate modeling of the physics behind the radiation-matter interaction processes, thus improving the dose determination in situations that are difficult for other algorithms ({\it e.~g.}, inhomogeneities and tissue interfaces). Particularly interesting is the case of small irradiation fields in which the charged particle equilibrium is lost, a problem that has gained importance due to the increasing incidence of intensity-modulated radiation therapy (IMRT). This treatment modality requires independent methods for quality assurance and therein MC algorithms may play a crucial role.

Despite the advantages of MC simulations, they have been mainly devoted to the verification of treatments planned with other algorithms. However, the increasing calculation power of the new computers has opened the possibility of computing treatment plans using MC algorithms to simulate both the linear accelerator and the computerized tomography image of the patient in times that begin to be acceptable in the routine clinical practice. There is, however, a drawback with MC dosimetry. While MC codes usually produce doses in units of eV/g per primary particle, those used in clinical dosimetry are expressed in Gy per monitor unit (MU). A conversion factor between these two units could be found if the current intensity of the electron beam impinging the target is known.

As the current is usually unknown, there are several procedures aiming at estimating the MC absolute dose (Francescon {\it et al.} 2000, Popescu {\it et al.} 2005, Lax {\it et al.} 2006, Panettieri {\it et al.} 2007). In general, all of them are based on the determination of the ratio between the MC dose at a reference point in a water phantom and the output in Gy/MU given by the calculation algorithm of the treatment planning system used (Francescon {\it et al.} 2000, Lax {\it et al.} 2006, Panettieri {\it et al.} 2007). A more accurate approach is that developed by Popescu {\it et al.} (2005) in which the dose delivered to the monitor ionization chamber of the linac is also taken into account. In principle, these absolute dosimetry formalisms can be applied to any radiation field configuration, open or blocked, small or large, in- or off-axis.

In the present work we apply these methodologies to small off-axis radiation fields. In particular, we consider the radiation fields currently used at the University Hospital of Essen for the treatment of retinoblastoma, the most common intraocular malignancy in early childhood. The Essen procedure (Sauerwein and Stannard 2009) is a technique developed from that originally 
proposed by Schipper (Schipper 1983, Schipper {\it et al.} 1997). Specifically, a dedicated `D'-shaped collimator, inserted in the accessory tray holder of a Varian Clinac~2100~C/D, operating at 6~MV, is used. This collimator can conform irradiation fields to two different sizes, 3.1~cm$^2$ or 5.2~cm$^2$, if an optional brass insert is included or not, respectively. The purpose of the collimator is twofold: (i) to reduce the absorbed dose to normal tissues, thus lowering the incidence of radiation-induced secondary tumors, and (ii) to reach a better definition of the beam penumbrae, thus improving the lens protection. An accurate dosimetry in the globe and the adnexa is necessary (Fl\"uhs {\it et al.} 1997, ICRU 2004, Schueler {\it et al.} 2006). However, it is complicated to attain and MC simulation appears to be a good tool to tackle it. In fact, MC simulation has permitted to accurately describe the dose distribution in small-sized irradiated volumes of the eyes (Thomson {\it et al.} 2008, Brualla {\it et al.} 2012b, Chiu {\it et al.} 2012, Miras {\it et al.} 2013).

In previous works, and using the MC code {\sc penelope} (Bar\'o {\it et al.} 1995, Sempau {\it et al.} 1997, Salvat {\it et al.} 2011), the relative absorbed dose distribution produced by the aforementioned `D'-shaped collimator in a water phantom was analyzed in detail (Brualla {\it et al.} 2012a, Mayorga {\it et al.} 2014). In this work we aim at evaluating the ability of the aforementioned procedures for determining the MC absolute dose in the case of this collimator and comparing the results to experimental data and to the dose distribution obtained with the analytical anisotropic algorithm (AAA) (Ulmer and Harder 1995, Ulmer and Kaissl 2003, Sievinen {\it et al.} 2005, VanEsch {\it et al.} 2006).

\section{Material and methods}

\subsection{Determination of the absolute dose}

In the approach of Popescu {\it et al.} (2005) the MC total absolute dose per monitor unit (MU) deposited in the phantom at the position $(x,y,z)$ is calculated as
\beq
D_{\rm MC}(x,y,z) \, = \, d(x,y,z) \, \frac{d_{\rm ch}^{\rm ref}}{d_{\rm ch}} \, \frac{D_{\rm cal}^{\rm ref}}{d_{\rm cal}^{\rm ref}} \, .
\label{eq:Popescu}
\eeq
Here $d(x,y,z)$ is the dose per primary electron, deposited in the phantom, in a scoring voxel that is centered at the position $(x,y,z)$. A ``primary electron'' refers to an electron emitted from the source that impinges on the linac target. In equation (\ref{eq:Popescu}), $d_{\rm ch}^{\rm ref}$ and $d_{\rm ch}$ are doses per primary electron deposited in the monitor chamber included in the simulation geometry of the linac. It is assumed that these doses are average values scored in the whole active volume of the monitor chamber. The superscript ``ref'' indicates that the dose must be obtained in reference conditions, that is for a field size of 10~cm~$\times$~10~cm and a source-to-surface 
distance (SSD) of $90\,$cm. $D_{\rm cal}^{\rm ref}$ is the dose per monitor unit, in reference conditions, at the calibration point, which is commonly situated at the isocenter, at a depth of 10 cm in the water phantom. Finally, $d_{\rm cal}^{\rm ref}$ is the dose per primary electron, deposited in a scoring voxel that is centered at the calibration point, in reference conditions.
The lower case doses $d$ appearing in the r.h.s. of equation (\ref{eq:Popescu}) are obtained in MC simulations while $D_{\rm cal}^{\rm ref}$ is an experimental value of the specific linac for which the analysis is carried out.

The dose scored in the monitor chamber can be separated in two terms as follows:
\beq
d_{\rm ch} \, = \, d_{\rm ch,f} \, + \, d_{\rm ch,b} \, .
\label{eq:chamber}
\eeq
Here $d_{\rm ch,f}$ is the contribution due to the beam particles that enter the monitor chamber from above, following the beam incident flux, and $d_{\rm ch,b}$ is that of those particles that enter the monitor chamber after being backscattered on the jaws, and the remaining elements of the linac head geometry situated downstream below the chamber. For a given operation energy of the linac, $d_{\rm ch,f}$ is the same for all configurations. 
Conversely, $d_{\rm ch,b}$ may depend on the field aperture defined by the collimation system of the linac. Then, the total dose in equation (\ref{eq:Popescu}) can be written as

\beq
D_{\rm MC}(x,y,z) \, = \, d(x,y,z) \, R_{\rm ch} \,
D_{\rm cal}^{\rm ref} \, ,
\label{eq:Pop3}
\eeq
where
\beq
R_{\rm ch} \, = \, \frac{S_{\rm ch}}{d_{\rm cal}^{\rm ref}}  \, .
\label{eq:Rch}
\eeq
Here
\beq
S_{\rm ch} \, = \, \frac{d_{\rm ch,f}\,+\,d_{\rm ch,b}^{\rm ref}}{d_{\rm ch,f}\,+\,d_{\rm ch,b}} 
\label{eq:Sb-factor}
\eeq
is the so-called monitor chamber backscatter factor. Recently, Zavgorodni {\it et al.} (2014) combined measured $S_{\rm ch}$ factors with simulated $d_{\rm ch,b}^{\rm ref}$ and $d_{\rm ch,b}$  (which are rather insensitive to the geometrical details of the monitor chamber) to determine, through equation (\ref{eq:Sb-factor}), the dose $d_{\rm ch,f}$ to be used in equations (\ref{eq:Pop3})-(\ref{eq:Sb-factor}) for various qualities of a 21EX and a TrueBeam Varian linac. 

Other authors (Francescon {\it et al.} 2000, Lax {\it et al.} 2006, Panettieri {\it et al.} 2007) proposed to calculate  the MC total dose per MU using a simpler expression:
\beq
D^{\rm app}_{\rm MC}(x,y,z) \, = \, d(x,y,z) \,  \frac{D_{\rm cal}^{\rm ref}}{d_{\rm cal}^{\rm ref}} \, .
\label{eq:Francescon}
\eeq
This is equivalent to assume 
\beq
S_{\rm ch} \, \approx \, 1 
\label{eq:chamber-app}
\eeq
in the prescription of Popescu {\it et al.}

\subsection{`D'-shaped collimation system}
\label{sec:collimator}

\begin{figure}
\centering
\includegraphics[width=11.cm]{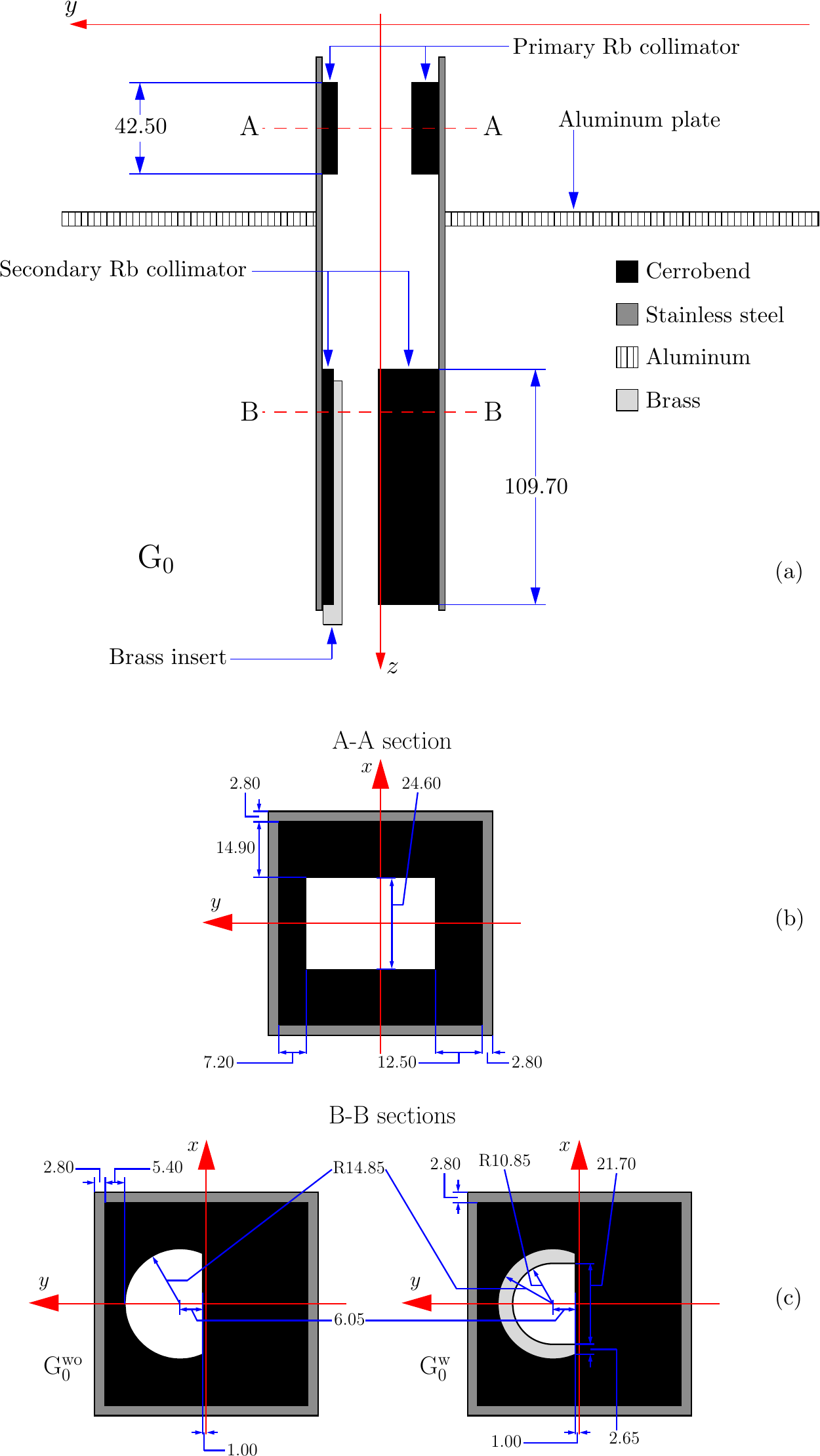}
\caption{Scheme of the geometry of the dedicated `D'-shaped collimator considered in the present work. Panel (a) shows the $x=0$ cut of the geometry of the whole collimator. Panel (b) and (c) represent the transverse A-A and B-B sections indicated in panel (a). The left (c)  panel corresponds to the largest irradiation field of 5.2~cm$^2$ while the right one shows the geometry when the optional brass insert is included and the field reduces to 3.1~cm$^2$. Dimensions are inmm. The blueprint of this collimator can be found in the work by Brualla {\it et al.} (2012a).
\label{fig:G0}
}
\end{figure}

As indicated above, the aim of the present work is to apply the approaches described in the previous subsection to the irradiation fields provided by a `D'-shaped collimation system used for the retinoblastoma treatment at the University Hospital of Essen. The whole collimation system is described in detail in the work by Brualla {\it et al.} (2012a). Its transverse sections that show up the shape of the two irradiation fields available are shown in figure~\ref{fig:G0}. The collimator is made of Cerrobend and patients can be treated with either
3.1~cm$^2$ (right panel) or 5.2~cm$^2$ (left panel) fields according to whether the insert is present or not. The configuration without the brass insert is referred as G$_0^{\rm wo}$ geometry, while G$_0^{\rm w}$ labels that including it. 

The collimator is inserted in the accessory tray holder of a Varian Clinac~2100~C/D operating at 6~MV. Recently (Brualla {\it et al.} 2012a), the dosimetry of this system was analyzed by comparing results obtained in MC simulations with {\sc penelope}, experimental measurements in a water phantom and the predictions of the AAA implemented in the Eclipse (Varian Medical Systems, Palo Alto, California) treatment planning system. Also the dosimetric role of the various elements included in the collimation system was analyzed with MC simulations (Mayorga {\it et al.} 2014).

In figure~\ref{fig:G0} the coordinate system used in all calculations is indicated. The direction of increasing $z$ coincides with the beam central axis. The origin of coordinates is situated upstream, at $100$~cm from the isocenter. At the isocenter, the radiation field defined by the jaws is centered at $x = 0$ and $y = 1.4$~cm. It is symmetric, with $5.5\,$cm width, about the $y$ axis as defined by the $x$-jaws of the linac. In the $y$-direction, the $y$-jaws open $0.7\,$cm in the negative $y$-axis and $3.5\,$cm in the positive one, defining a field $4.2\,$cm wide. The movable jaws are situated in the same position independently whether the brass insert is used or not.

\subsection{Simulation details}

As said above, all $d$-doses in equations (\ref{eq:Popescu})--(\ref{eq:Sb-factor}) were obtained from MC simulations in which the complete geometry of the linac head and the collimation system were taken into account. Simulations were carried out with {\sc penEasy} (Sempau {\it et al.} 2011), a main steering code that uses the general-purpose MC system {\sc penelope}. 

{\sc penelope} simulates the coupled transport of electrons, positrons and photons. This MC system performs the simulations using a mixed scheme in which electron and positron interactions are classified as hard or soft events. In hard events the angular deflections and/or the energy losses are larger than certain cutoffs; these events are simulated in a detailed way. All soft interactions occurring between two hard events are described within a multiple scattering theory; in particular, their simulation is done in terms of a single artificial event. Photons are simulated on a detailed scheme, that is simulating all their interactions in chronological order. Particle transport is carried out until the particle kinetic energy is below user-defined absorption energies ($E_{\rm abs}$), and the particle is locally absorbed. The multiple scattering algorithm is controlled by the following parameters: $C_{1}$ is related to the average angular deflection due to a hard elastic collision and all previous soft collisions. $C_{2}$ controls the maximum allowed value for the average fractional energy loss in a step. $W_{\rm CC}$ and $W_{\rm CR}$ are energy cutoffs for hard inelastic collisions and hard bremsstrahlung emission, respectively. If the user sets $C_{1} = C_{2} = 0$ and $W_{\rm CR}$ equal to an arbitrary negative number, {\sc penelope} performs a detailed simulation for electrons and positrons.

The whole detailed head of a Varian Clinac~2100~C/D operating at 6~MV, the retinoblastoma collimator and a water phantom of $40\,{\rm cm}\times 40\,{\rm cm}\times 40\,{\rm cm}$ were included in the simulations.
The linac geometry was generated with the code {\sc penEasyLinac} (Brualla {\it et al.} 2009a, Sempau {\it et al.} 2011, Rodriguez {\it et al.} 2013). Specific benchmarks of the calculations with this set of codes for the linac and nominal energy considered in this work  were done in previous publications (Fern\'andez-Varea {\it et al.} 2007, Brualla {\it et al.} 2009a, Brualla {\it et al.} 2009b, Panettieri {\it et al.} 2009, Sempau {\it et al.} 2011)

Simulations were performed in several steps in which the various $d$-doses were calculated. 
A detailed description of these steps can be found in the Appendix. In the simulations performed the following variance-reduction techniques (VRTs) were applied: Movable-skins (Brualla {\it et al.} 2009b) were used in the linac primary collimator and the jaws. Interaction forcing (Salvat {\it et al.} 2011) was applied to bremsstrahlung interactions in the photon target. Rotational splitting (Brualla and Sauerwein 2010) was used in the upper part of the linac, above the jaws, where cylindrical symmetry holds. Standard particle splitting was used in the water phantom.
 
In a recent work Rodriguez {\it et al.} (2015) pointed out the disagreement between simulated and experimental dose distributions obtained from a linac when a long step length was used for the simulation of the bremsstrahlung emission in the target. Specifically, using values of $C_{1}$ and $C_{2}$ greater than 10$^{-3}$ for the tungsten of the target may produce a bias in the bremsstrahlung distribution that otherwise disappears almost completely when values of 10$^{-3}$ or smaller are used for both tracking parameters. In order to investigate the effect of this new set of tracking parameters on absolute dosimetry in Gy/MU, all simulations were run with the tracking parameters for the target indicated in table \ref{tab:param}, namely $C_{1}=C_{2}=0.1$, $W_{\rm CC}=100\,$keV and $W_{\rm CR}=20\,$keV, and with the parameter set proposed by Rodriguez {\it et al.} (2015) for the tungsten of the target: $C_{1}=C_{2}=0.001$, $W_{\rm CC}=1\,$keV and $W_{\rm CR}=20\,$keV. The results obtained with both sets of parameters were labeled P$_1$ and P$_2$, respectively. 

\subsection{Comparison to experimental measurements}

In order to test the validity of the approaches described, experimental doses measured at the point $(x_0=0,\,y_0=1.03\,{\rm cm},\,z_0=100\,{\rm cm})$ were compared to the MC doses given by equation (\ref{eq:Pop3}). This point coincides with the center of curvature of the circular part of the field at the measurement plane ($z_0=100\,{\rm cm}$).
Actually, the comparison was done as follows. According to equation (\ref{eq:Pop3}) we can write
\beq
{\cal D}_{\rm MC}(x_0,y_0,z_0)\, = \, \frac{D_{\rm MC}(x_0,y_0,z_0)}{D_{\rm cal}^{\rm ref}} \, = \, d(x_0,y_0,z_0) \, R_{\rm ch} \,  .
\label{eq:Pop4}
\eeq
Similarly, for the simpler approach one gets
\beq
{\cal D}^{\rm app}_{\rm MC}(x_0,y_0,z_0)\, = \, \frac{D^{\rm app}_{\rm MC}(x_0,y_0,z_0)}{D_{\rm cal}^{\rm ref}} \, = \, d(x_0,y_0,z_0) \, \frac{1}{d_{\rm cal}^{\rm ref}} \,  .
\label{eq:Fra4}
\eeq
These normalized doses ${\cal D}_{\rm MC}$ are dimensionless quantities that were compared to the corresponding experimental ones given by
\beq
{\cal D}_{\rm exp}(x_0,y_0,z_0) \, = \, \frac{D_{\rm exp}(x_0,y_0,z_0)}{D_{\rm cal}^{\rm ref}} \, .
\label{eq:calD}
\eeq

Measurements were carried out according to the protocols DIN~6800-2 (DIN 2008), for the reference doses, and DIN~6809-8 (DIN 2014), for the small field doses. The size of the alignment field was $5\times 5\,$cm$^2$. The experimental doses were measured with a PTW M31016 ionization chamber, which has an active volume of $0.016\,$cm$^{3}$, at the point $(x_0,y_0,z_0)$ with an isocentric configuration. In this way it was possible to obtain measurements at different depths by varying the SSD from 96 to $99\,$cm in $1\,$cm step. The two configurations for the retinoblastoma collimator, G$_0^{\rm wo}$ and G$_0^{\rm w}$, were considered. The values selected for $x_0$ and $y_0$ permit to maximize the dose rate in the treatment field of the retinoblastoma. As we had five experimental determinations of each dose $D_{\rm exp}(x _0,y_0,z_0)$, taken during the last four years, we compared the MC values to the corresponding averages over the five experimental values, $\overline{\cal D}_{\rm exp}(x_0,y_0,z_0)$. Also the corresponding calibration dose $D_{\rm cal}^{\rm ref}$ was measured in reference conditions.

The doses $d(x_0,y_0,z_0)$ required to obtain the MC normalized doses were calculated by considering a scoring voxel of 0.016 cm$^{3}$ centered around the same point $(x_0, y_0, z_0)$ where the effective point of measurement of the chamber was situated and with the same geometrical conditions for the phantom positioning. Results for both P$_1$ and P$_2$ parameter sets were determined. 

Statistical uncertainties with a coverage factor $k=1$ were estimated for all quantities considered. In the case of the MC doses, these uncertainties were obtained directly from the corresponding simulations. For the normalized doses the linear propagation prescription was assumed.

In order to complete the analysis, the average experimental values $\overline{\cal D}_{\rm exp}(x_0,y_0,z_0)$ were also compared to 
\beq
{\cal D}_{\rm AAA}(x_0,y_0,z_0)\, = \, \frac{D_{\rm AAA}(x_0,y_0,z_0)}{D_{\rm cal}^{\rm ref}} \, ,
\label{eq:FAAA}
\eeq
where $D_{\rm AAA}(x,y,z)$ are the doses obtained using the Eclipse treatment planing system that implements the AAA as calculation engine. Eclipse (v. 8.9.09) on ARIA 8 with AAA (v. 8.9.08) were used. In Eclipse, the detailed geometry of the retinoblastoma collimator cannot be simulated and calculations were done by considering a block $10\,$cm thick, with 0.1\% transmission, that includes an aperture with 100\% transmission. The shape of this aperture is designed according to the two fields defined by the retinoblastoma collimator, including or not the brass insert. A $0.1\,$cm calculation grid size was used and scoring voxels of $0.086\times 0.086\times 0.2\,$cm$^3$ were defined in the water phantom. 

\section{Results and discussion}

The doses $d_{\rm cal}^{\rm ref}$, $d_{\rm ch,f}$ and $d_{\rm ch,b}^{\rm ref}$ obtained in the simulations performed with the two parameter sets P$_1$ and P$_2$ are given in table \ref{table:dosisCIrtr}. The calculation of the backward dose $d_{\rm ch,b}$ does not show differences between the two geometries analyzed G$_{0}^{\rm wo}$ and G$_{0}^{\rm w}$, therefore only one value is reported. This is due to the fact that both geometries share the same position of the jaws, which produce the largest contribution to the backscatter radiation in the ionization chamber. Also, $R_{\rm ch}$ and $S_{\rm ch}$ are the same for these two geometries.

The change in the tracking parameters produces a noticeable increase of 12.6\% in $d_{\rm cal}^{\rm ref}$. In contrast, the forward dose in the chamber is slightly reduced while the backward ones are increased by a few percent. As the factor $R_{\rm ch,cal}^{\rm ref}$, given in equation (\ref{eq:Rch}), is inversely proportional to the dose $d_{\rm cal}^{\rm ref}$, the values of this factor found for P$_1$ are $\sim 10\%$ larger than those obtained with the P$_2$ parameter set.

\begin{table}[!t]
\begin{center}
\begin{tabular}{cccc}
\hline \hline 
 & P$_1$ & P$_2$ & relative difference \\
\hline
$d_{\rm cal}^{\rm ref}$  & 0.635(3) & 0.715(3) & 12.6\% \\ 
$d_{\rm ch,f}$ & $12.7310(7)$ & $12.6400(4)$ & -0.7\% \\ 
$d_{\rm ch,b}^{\rm ref}$ & $1.6697(2) \cdot 10^{-2}$ & $1.7315(2) \cdot 10^{-2}$ & 3.7\% \\ 
$d_{\rm ch,b}$ & $4.6225(9) \cdot 10^{-3}$ & $4.8089(9) \cdot 10^{-3}$ & 4.0\% \\
$R_{\rm ch}$ & 1.576(8) & 1.399(7) & -11.2\% \\ 
$S_{\rm ch}$ & 1.00095(8) &  1.00099(5)   & 0.41\%\\     
\hline \hline 
\end{tabular}
\end{center}
\vspace*{-0.5cm}
\caption{\label{table:dosisCIrtr}
Doses, expressed in eV/g per primary particle, in the monitor chamber for the reference conditions and for the geometries analyzed, obtained with the two parameter sets considered. Derived quantities $R_{\rm ch}$, as given by equation (\ref{eq:Rch}), and $S_{\rm ch}$, as given by equation (\ref{eq:Sb-factor}), are shown. 
Uncertainties, with a coverage factor $k=1$ are given between parentheses; {\it e. g.},  1.7315(2) indicates $1.7315\pm 0.0002$.}
\end{table}

\begin{table}[!b]
\begin{center}
{\scriptsize
\begin{tabular}{cccccccccccc}
\hline 
\hline 
&&& \multicolumn{9}{c}{$D_{\rm exp}(x_0=0, y_0=1.03\,{\rm cm}, z_0=100\,{\rm cm})$ (Gy/100\,{\rm MU})} \\ \cline{4-12}
&&& \multicolumn{4}{c}{G$_{0}^{\rm wo}$}&&\multicolumn{4}{c}{G$_{0}^{\rm w}$} \\ \cline{4-7} \cline{9-12}
set &$D_{\rm cal}^{\rm ref}$ (Gy/100\,{\rm MU})&~~&SSD=99$\,{\rm cm}$ & SSD=98$\,{\rm cm}$ & SSD=97$\,{\rm cm}$ & SSD=96$\,{\rm cm}$ 
&~~&SSD=99$\,{\rm cm}$ & SSD=98$\,{\rm cm}$ & SSD=97$\,{\rm cm}$ & SSD=96$\,{\rm cm}$ \\ \hline
1 & 0.808 && 0.943 &0.972 &0.938 &0.900 &&0.938 &0.961 &0.927 &0.886\\
2 & 0.802 &&0.923 &0.952 &0.918 &0.881 &&0.924 &0.951 &0.915 &0.877\\
3 & 0.805 && 0.932 &0.947 &0.914 &0.876 &&0.929 &0.945 &0.908 &0.869\\
4 & 0.804 &&0.933 &0.954 &0.920 &0.885 &&0.929 &0.950 &0.916 &0.879\\
5 & 0.803 &&0.942 &0.969 &0.927 &0.890 &&0.936 &0.955 &0.921 &0.884\\ 
\hline
\multicolumn{3}{r}{$\overline{\cal D}_{\rm exp}(x_0,y_0,z_0)$} & 1.162(9) & 1.192(9) & 1.148(9) & 1.102(8) && 1.158(9) & 1.184(9) & 1.140(9) & 1.093(9)\\
\hline\hline 
\end{tabular}
}
\end{center}
\vspace*{-0.5cm}
\caption{\label{tab:Dexp}Experimental doses, in Gy$/100\,$MU, for the five measurement sets available. The calibration dose obtained in reference conditions $D_{\rm cal}^{\rm ref}$ and the doses $D_{\rm exp}(x_0,y_0,z_0)$, used to test the validity of the procedure described in the present work and found for the two geometries analyzed, G$_{0}^{\rm wo}$ y G$_{0}^{\rm w}$, at four different SSDs, are given. Relative uncertainties of these measurements are 1.3\%. Last row represents the average values of the experimental normalized doses given by equation (\ref{eq:calD}), for each SSD. Uncertainties, with a coverage factor $k=1$ are given between parentheses; {\it e.~g.},  1.152(9) indicates $1.152\pm 0.009$.}
\end{table}

\begin{table}[!t]
\begin{center}
{\scriptsize
\begin{tabular}{ccccccccccc}
\hline 
\hline 
&& \multicolumn{9}{c}{$d(x_0=0, y_0=1.03\,{\rm cm}, z_0=100\,{\rm cm})$ (eV/g/particle)} \\ \cline{3-11}
&& \multicolumn{4}{c}{P$_1$}&&\multicolumn{4}{c}{P$_2$} \\ \cline{3-6} \cline{8-11}
 &~~&SSD=99$\,{\rm cm}$ & SSD=98$\,{\rm cm}$ & SSD=97$\,{\rm cm}$ & SSD=96$\,{\rm cm}$ 
&~~&SSD=99$\,{\rm cm}$ & SSD=98$\,{\rm cm}$ & SSD=97$\,{\rm cm}$ & SSD=96$\,{\rm cm}$ \\ \hline
G$_{0}^{\rm wo}$ && 0.746(9) & 0.775(9) & 0.746(8) & 0.716(7) && 0.830(9) & 0.862(8) & 0.834(7) & 0.801(7) \\
G$_{0}^{\rm w}$  && 0.740(8) & 0.765(7) & 0.739(7) & 0.708(7) && 0.825(9) & 0.856(8) & 0.824(7) & 0.789(7) \\
\hline\hline 
\end{tabular}
}
\end{center}
\vspace*{-0.5cm}
\caption{\label{tab:d-sim}Doses $d(x_0,y_0,z_0)$, in eV/g per primary particle, obtained in the simulations for the three geometries analyzed in the present work, G$_{0}^{\rm wo}$ and G$_{0}^{\rm w}$, at the four different SSDs considered. Uncertainties, with a coverage factor $k=1$ are given between parentheses; {\it e.~g.}, 0.746(9) means $0.746\pm 0.009$.}
\end{table}

We also analyzed the effect produced by the approximation (\ref{eq:chamber-app}) in the absolute dose calculation. The values of the ratio $S_{\rm ch}=d_{\rm ch}^{\rm ref}/d_{\rm ch}$ are given in table \ref{table:dosisCIrtr}. The difference found between the results for both parameter sets are well below 1\% and the correction introduced when the ratio of equation~(\ref{eq:Sb-factor}) is considered is 0.1\% at most.

Table \ref{tab:Dexp} summarizes the experimental data used in our comparison. The doses found for the five different sets are shown for the four SSD values and the two collimator configurations at $x_0=0,\,y_0=1.03\,{\rm cm},\,z_0=100\,{\rm cm}$. Also the corresponding calibration doses $D_{\rm cal}^{\rm ref}$ are given (second column). The average normalized doses  $\overline{\cal D}_{\rm exp}(x_0,y_0,z_0)$ are shown in the last row.

Table \ref{tab:d-sim} shows the doses $d(x_0,y_0,z_0)$ obtained in the various simulations carried out with the two configurations of the retinoblastoma collimator studied in the present work. The values for the two parameter sets P$_1$ and P$_2$ are given. Those corresponding to P$_2$ are larger than those of P$_1$ and the differences are slightly above 10\% at most. These differences, together with those found in some of the dose values tabulated in table \ref{table:dosisCIrtr}, point out the relevant effect produced by the modification in the tracking parameters considered in the step 1 of the linac simulation process (see Appendix).

\begin{figure}[!b]
\begin{center}
\includegraphics[width=8cm]{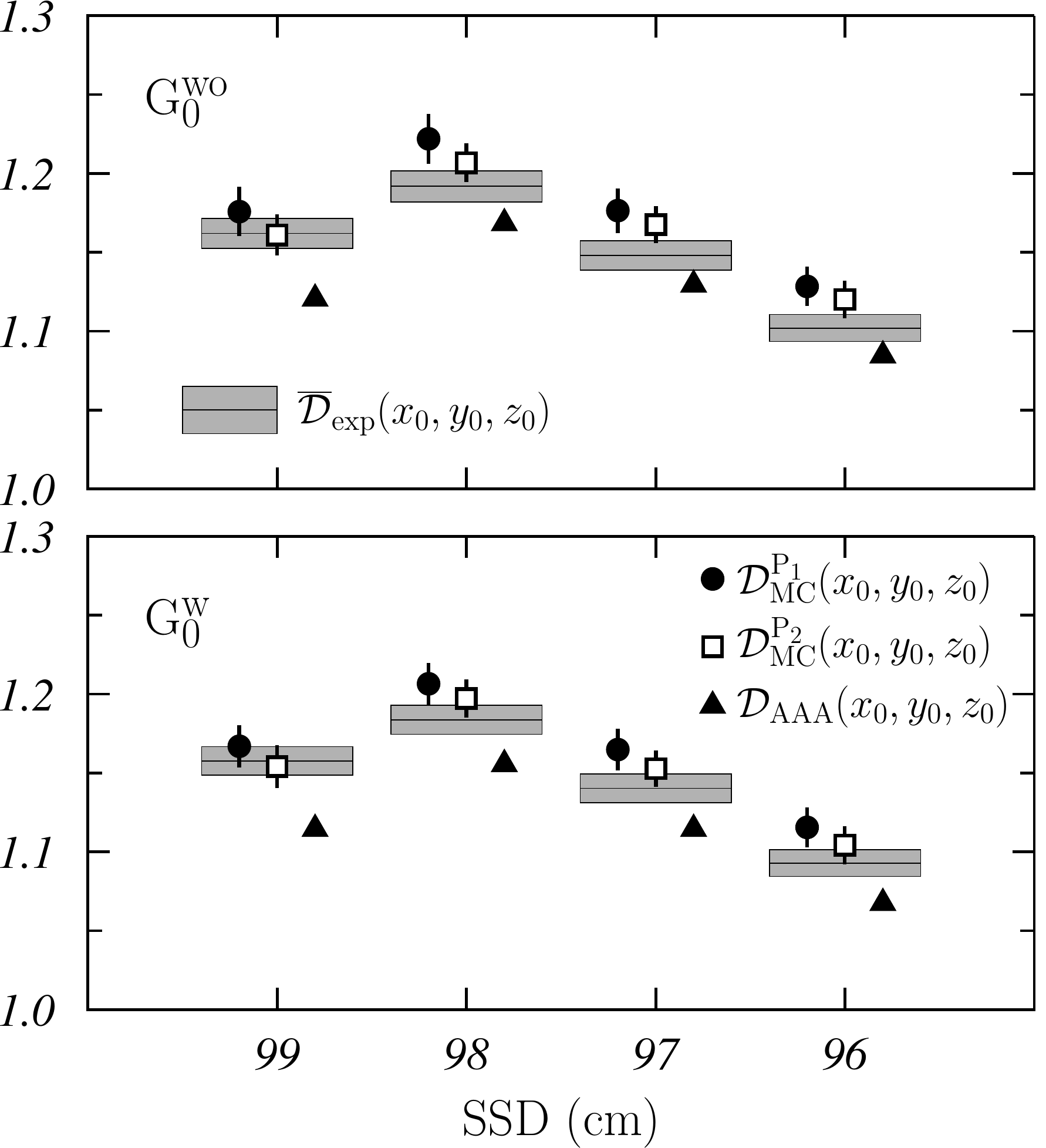}
\end{center}
\vspace*{-0.5cm}
\caption{\label{fig:doses}Comparison between the average experimental ratios $\overline{\cal D}_{\rm exp}(x_0,y_0,z_0)$ (gray boxes), the values of ${\cal D}^{{\rm P}_i}_{\rm MC}(x_0,y_0,z_0)$ (see equation (\ref{eq:Pop4})), obtained with the two tracking parameters sets P$_1$ (solid circles) and P$_2$ (open squares), and the values ${\cal D}_{\rm AAA}(x_0,y_0,z_0)$ (see equation (\ref{eq:FAAA})), obtained with AAA, for the four SSDs considered. Uncertainties are given with a coverage factor $k=1$.Uncertainties of the experimental data are represented by the height of the gray boxes.}
\end{figure}

The results obtained using the simulated doses are in agreement, within the uncertainties, with the experimental ones for P$_2$ while for P$_1$ this is true only if a coverage factor $k=2$ is considered. This can be seen in a clearer manner in figure \ref{fig:doses} where the normalized experimental doses $\overline{\cal D}_{\rm exp}(x_0,y_0,z_0)$ (gray rectangles) are compared to the normalized simulated ${\cal D}^{{\rm P}_i}_{\rm MC}(x_0,y_0,z_0)$ (solid circles for P$_1$ and open squares for P$_2$)  for G$_{0}^{\rm wo}$ (upper panel) and G$_{0}^{\rm w}$ (lower panel).

Regarding the comparison between simulations, it is worth mentioning  that the doses ${\cal D}^{\rm app}_{\rm MC}(x_0,y_0,z_0)$, given by equation (\ref{eq:Fra4}), are statistically compatible with the corresponding ${\cal D}_{\rm MC}(x_0,y_0,z_0)$. In fact ${\cal D}_{\rm MC}/{\cal D}^{\rm app}_{\rm MC}=S_{\rm ch}$, which amounts to
$1.00095(8)$ for P$_1$ and 1.00099(5) for P$_2$ (see table~\ref{table:dosisCIrtr}).

The relative differences between experimental and calculated doses 
are below 2.6\% for P$_1$ and 1.7\% for P$_2$. These maximum differences are around the maximum value of 2\% found by Popescu {\it et al.} (2005). These authors used the BEAMnrc/DOSXYZnrc codes (Rogers {\it et al.} 2001,Walters and Rogers 2003) to simulate both open and blocked fields, including those employed in IMRT.
It is worth pointing out that the differences with the experimental values are larger for the geometry G$_{0}^{\rm wo}$ than for G$_{0}^{\rm w}$, when the brass insert is present and the irradiation field is smaller: in this last case the maximum differences are below 2.2\% and 1.2\% for P$_1$ and P$_2$, respectively.
In general, the use of the tracking parameter set P$_2$ improves in all cases analyzed the agreement with the experimental data.

The doses ${\cal D}^{{\rm P}_2}_{\rm MC}(x_0,y_0,z_0)$ are always smaller than the corresponding ${\cal D}^{{\rm P}_1}_{\rm MC}(x_0,y_0,z_0)$. Differences between the results obtained with the two tracking parameter sets are below 1.3\% in absolute value.

Figure \ref{fig:doses} also includes the 
normalized doses ${\cal D}_{\rm AAA}(x_0,y_0,z_0)$, given by equation (\ref{eq:FAAA}), which are shown by the triangles. 
AAA underestimates the experimental doses in all cases calculated and shows a 
better agreement with the experiment in the case of G$^{\rm wo}_0$ than for G$^{\rm w}_0$. 

At ${\rm SSD}=99\,$cm, the relative differences between AAA and the experiment are larger than 3\% in absolute value. This contrasts with the fact that the doses obtained for both ${\rm P}_1$ and ${\rm P}_2$ show the 
best agreement with the experiment precisely for this SSD.
For the other SSDs,
the relative differences between AAA results and the experimental doses 
are similar, in magnitude, to those corresponding to P$_2$ in the case of G$^{\rm wo}_0$, while for G$^{\rm w}_0$ are much larger, above 2\%, that is of the same order of the relative differences found for P$_1$.
This points out that the accuracy of AAA decreases when the field size reduces. Similar findings have been quoted by other groups (see {\it e. g.}, Ong {\it et al.} 2011).

\section{Conclusions}

Two formalisms proposed to estimate total absolute doses from MC simulations have been evaluated for the case of small off-axis fields defined by a collimator used in the retinoblastoma treatment with external beams produced by a linac.

The simulation of the linac has been carried out with {\sc penelope} and its geometry has been generated with {\sc penEasyLinac}. Two tracking parameters sets, P$_1$ and P$_2$, have been considered for the simulation of the particle transport in the target of the linac, the second one characterized by extremely low values of $C_1$, $C_2$ and $W_{\rm CC}$ and providing a nearly detailed simulation of charged particles in the target. The change in the parameters produced non-negligible differences, above 10\%, in doses tallied in reference conditions.

The absolute doses found using the procedure of Popescu {\it et al.} have been compared to measured doses for different SSDs. The MC results differ from the experimental ones by 2.6\% and 1.7\% for P$_1$ and P$_2$, respectively. The absolute doses obtained for the two parameter sets in the two geometries analyzed show differences below 1.3\%, despite the aforementioned disagreement in $d_{\rm cal}^{\rm ref}$. These results validate the use of the approach proposed by Popescu and collaborators in the case of the small, `D'-shaped and off-axis radiation fields considered herein. The approximation $d_{\rm ch}^{\rm ref}/d_{\rm ch}\approx 1$ amounts to an error below 0.1\% and can be used in the studied cases.

\section*{Acknowledgements}
 
The work of PAM and AML has been supported in part by the Junta de
Andaluc\'{\i}a (FQM0220,FQM0387) and the Spanish Ministerio de Econom\'{\i}a y Competitividad and FEDER (projects FPA2012-31993 and FPA2015-67694). LB and WS are grateful to the Deutsche Forschungsgemeinschaft (project BR~4043/3-1).

\section*{Apendix. Details of the simulations}

Simulations were performed in several steps in which the various $d$-doses entering equations (\ref{eq:Popescu})-(\ref{eq:chamber-app}) were calculated. 
The values of the tracking parameters considered in the different simulation steps are shown in table \ref{tab:param}. In what follows a description of the main characteristics of each simulation step is given.

\begin{table}[!t]
\begin{center}
\begin{tabular}{cccccccccl}
\hline \hline 
&& \multicolumn{3}{c}{$E_{\rm abs}$} &&&&&\\ \cline{3-5}
step & material & $e^-$ & $\gamma$ & $e^+$ & $C_{1}$ & $C_{2}$ & $W_{\rm CC}$ & $W_{\rm CR}$ & Part of the geometry \\
\hline 
1 &Tungsten & $100$ & $20$ & $100$ & 0.1 & 0.1 & $100$ & $20$ & target [P$_1$]\\
 & & {\it 100} & {\it 20} & {\it 100} & {\it 0.001} & {\it 0.001} & ${\it 1}$ & {\it 20} & {\it target} [P$_2$] \\
 &Tungsten & $100$ & $20$ & $100$ & 0.1 & 0.1 & $100$ & $20$ & primary collimator (skins)\\
&Tungsten & $10^6$ & $100$ & $10^6$ & 0 & 0 & 0 & 0 & {primary collimator  (non-skins)}\\
&Air & $0.05$ & $0.05$ & $0.05$ & 0 & 0 & 0 & $-11$ & monitor chamber \\ 
&Air & $10^6$ & $10^6$ & $10^6$ & 0 & 0 & 0 & 0 & PSF$_{\rm sec}$\\ 
\hline
2 &Tungsten & $100$ & $20$ & $100$ & 0.1 & 0.1 & $100$ & $20$ & jaws (skins)\\
&Tungsten & $10^6$ & $100$ & $10^6$ & 0 & 0 & 0 & 0 & {jaws (non-skins)}\\
&Air & $0.05$ & $0.05$ & $0.05$ & 0 & 0 & 0 & $-11$ & monitor chamber \\ 
& Air & $10^6$ & $10^6$ & $10^6$ & 0 & 0 & 0 & 0 & PSF$_{\rm jaw}^0$, PSF$_{\rm jaw}^{\rm ref}$\\
\hline 
$2^\prime$ &Tungsten & $100$ & $20$ & $100$ & 0.1 & 0.1 & $100$ & $20$ & jaws \\
&Air & $0.05$ & $0.05$ & $0.05$ & 0 & 0 & 0 & $-11$ & monitor chamber \\
\hline 
3 & Cerrobend & $100$ & $20$ & $100$ & 0.1 & 0.1 & $100$ & $20$ & Rb collimator\\
   &Air & $10^6$ & $10^6$ & $10^6$ & 0 & 0 & 0 & 0 & PSF$_{\rm Rb}^{\rm wo}$, PSF$_{\rm Rb}^{\rm w}$\\
\hline 
4,$4^\prime$ &Air & $100$ & $20$ & $100$ & 0.1 & 0.1 & $100$ & $20$ & \\
&Water & $100$ & $20$ & $100$ & 0.1 & 0.1 & $100$ & $20$ & phantom \\
\hline \hline 
\end{tabular}
\end{center}
\vspace*{-0.5cm}
\caption{\label{tab:param}
Tracking parameters used in the various steps of the simulations carried out. The values of $E_{\rm abs}$, $W_{\rm CC}$ and $W_{\rm CR}$ are given in keV. Figures in italics correspond to the simulation parameters used for the case P$_2$.}
\end{table}

\begin{description}
\item[Step 1.] A simulation was run from the electron source downstream to the exit of the upper segment of the linac, that is, just above its secondary collimator (lead shield) where a phase-space file, PSF$_{\rm sec}$, was tallied. The primary electron source was modeled as a monoenergetic point-like pencil beam with zero divergence. Electrons were emitted with an initial kinetic energy of 6.26 MeV. This set of initial beam parameters is the same for P$_1$ and P$_2$ simulations and reproduces well the absorbed dose distribution of the Clinac 2100~C/D used at the University Hospital of Essen. It must be noticed that this particular linac is not tuned to reproduce the Golden Beam Data Set provided by Varian for this type of linacs. To determine the PSF$_{\rm sec}$, a very thin air slab was defined just above the secondary collimator and all particles reaching it were stopped by imposing arbitrarily high absorption energies ($1\,$GeV). In this step the movable skins VRT, with a thickness of $5\,$mm, was considered in the primary collimator of the linac, interaction forcing was used in the target, with a forcing factor of 100, and rotational splitting (with a splitting factor of 15) was applied to particles reaching a thin air slab situated just after the monitor ionization chamber, previously to the secondary collimator of the linac. Apart from tallying the PSF$_{\rm sec}$, in this step $d_{\rm ch,f}$ was determined by scoring the dose in the air-cavity of the monitor chamber. A total of $10^9$ primary electrons were simulated.

\item[Step 2.] In the second step the simulation through the jaws was carried out and two phase-space files were scored. One of them, PSF$_{\rm jaw}^0$, corresponded to the geometry G$_0$. It is worth pointing out that the position of the jaws was the same if the optional brass insert was present or not. The second phase-space file, PSF$_{\rm jaw}^{\rm ref}$, was tallied with the jaws situated in reference conditions. In this step the PSF$_{\rm sec}$ was used as the particle source. The movable skins VRT, also with $5\,$mm thickness, as in the primary collimator, was applied in the jaws. The two PSFs were accumulated in a thin air slab situated just below the jaws. 

\item[Step 2$^\prime$.] The second step was repeated in order to determine the dose scored in the monitor chamber due to backscattering in the jaws (as well as in all the remaining elements of the collimation system of the linac). In these simulations the doses $d_{\rm ch,b}^0$ (for the geometries G$_0^{\rm wo}$ and G$_0^{\rm w}$) and $d_{\rm ch,b}^{\rm ref}$ (corresponding to the reference conditions) were calculated. The main difference with the simulations of step \#2 is that no VRTs were applied in the jaws to avoid any possible bias in the evaluated backscattering doses. No PSFs were scored in this step.
 
\item[Step 3.] In the third step particles emitted from the phase-space file PSF$_{\rm jaw}^0$ were transported through the corresponding retinoblastoma collimators and new phase-space files were scored in a thin air slab situated at their exit. We labeled them as PSF$_{\rm RB}^{\rm wo}$ and PSF$_{\rm RB}^{\rm w}$, according to the geometries used, G$_{0}^{\rm wo}$ and G$_{0}^{\rm w}$, respectively.  
 
\item[Step 4.] In the last simulation step, the phase-space files PSF$_{\rm RB}^{\rm wo}$ and PSF$_{\rm RB}^{\rm w}$ were used as sources of particles that were emitted towards a water phantom of $40 \times 40 \times 40$~cm$^3$ where the dose values $d(x,y,z)$ were determined.

\item[Step 4$^\prime$.] This simulation was carried out to determine $d_{\rm cal}^{\rm ref}$. The phase-space file, PSF$_{\rm jaw}^{\rm ref}$ scored in Step 2 was used as source of particles emitting towards the water phantom that was situated at a ${\rm SSD}=90\,$cm. The dose of interest was calculated following the prescription of Popescu {\it et al.} (2005) and a voxel of $9\,$mm$^{3}$ centered at the point $(0, 0, 10\,{\rm cm})$ (isocenter) was used to score the dose.
\end{description}

All materials not explicitly indicated in table \ref{tab:param} share the same parameters as tungsten in the skin regions. The parameter $s_{\rm MAX}$, defining the maximum step length of a particle trajectory, was fixed in each geometry body to one tenth of its characteristic thickness.


\section*{References}


\noindent 
Bar\'o J, Sempau J, Fern\'andez-Varea J M and Salvat F 1995 PENELOPE: An algorithm for Monte Carlo simulation of the penetration and energy loss of electrons and positrons in matter {\it Nucl. Instrum. Meth. Phys. Res. B} {\bf 100} 31-46\\[-2mm]

\noindent 
Brualla L, Mayorga P A, Fl\"uhs A, Lallena A M, Sempau J and Sauerwein W 2012a Retinoblastoma external beam photon irradiation with a special `D'-shaped collimator: a comparison between measurements, Monte Carlo simulation and treatment planning system calculation {\it Phys. Med. Biol.} {\bf 57} 7741-51\\[-2mm]

\noindent 
Brualla L, Palanco-Zamora R, Wittig A, Sempau J and Sauerwein W 2009a Comparison between PENELOPE and eMC in small electron fields {\it Phys. Med. Biol.} {\bf 54} 5469-81\\[-2mm]

\noindent 
Brualla L, Salvat F and Palanco-Zamora R 2009b Efficient Monte Carlo simulation for multileaf collimators using geometry-related variance-reduction techniques {\it Phys. Med. Biol.} {\bf 54} 4131-49\\[-2mm]

\noindent 
Brualla L and Sauerwein W 2010 On the efficiency of azimuthal and rotational splitting for Monte Carlo simulation of clinical linear accelerators,'' {\it Radiat. Phys. Chem.} {\bf 79} 929-32\\[-2mm]

\noindent 
Brualla L, Zaragoza F J, Sempau J, Wittig A and Sauerwein W 2012b Electron irradiation of conjunctival lymphoma-Monte Carlo simulation of the minute dose distribution and technique optimization {\it Int. J. Radiat. Oncol. Biol. Phys.} {\bf 83} 1330-7\\[-2mm]

\noindent 
Chiu-Tsao S-T, Astrahan M A, Finger P T, Followill D S, Meigooni A S, Melhus C S, Mourtada F, Napolitano M E, Nath R, Rivard M J, Rogers D W O and Thomson R M 2012 Dosimetry of $^{125}$I and $^{103}$Pd COMS eye plaques for intraocular tumors: Report of Task Group 129 by the AAPM and ABS {\it Med. Phys.} {\bf 39} 6161-84\\[-2mm]

\noindent 
DIN Deutsches Institut f\"ur Normung 2008 Dosismessverfahren nach der Sondenmethode f\"ur Photonen- und Elektronenstrahlung - Teil 2: Dosimetrie Hochenergetischer Photonen- und Elektronenstrahlung mit Ionisationskammern DIN 6800-2 (Belin: Deutsches Institut f\"ur Normung)\\[-2mm]

\noindent 
DIN Deutsches Institut f\"ur Normung 2014 Klinische Dosimetrie Teil 8: Dosimetrie Kleiner Photonen Bestrahlungsfelder DIN 6809-8 (Belin: Deutsches Institut f\"ur Normung)\\[-2mm]

\noindent 
Fern\'andez-Varea J M, Carrasco P, Panettieri V and Brualla L 2007 Monte Carlo based water/medium stopping-power ratios for various ICRP and ICRU tissues {\it Phys. Med. Biol.} {\bf 52} 6475-83\\[-2mm]

\noindent 
Fl\"uhs D, Bambynek M, Heintz M, Indenk\"ampen F, Kolanoski H, Wegener D, Sauerwein W and Quast~U 1997 Dosimetry and design of radioactive eye plaques {\it Front. Radiat. Ther. Oncol.} {\bf 30} 26-38 (1997).\\[-2mm]

\noindent 
Francescon P, Cavedon C, Reccanello S and Cora S 2000 Photon dose calculation of a three-dimensional treatment planning system compared to the Monte Carlo code BEAM {\it Med. Phys.} {\bf 27} 1579-87\\[-2mm]

\noindent 
International Commission on Radiation Units and Measurements 2004 Dosimetry of beta rays and low-energy photons for brachytherapy with sealed sources ICRU 72 (Bethesda: ICRU)\\[-2mm]

\noindent 
Lax I, Panettieri V, Wennberg B, Duch M A, N\"aslund I, Baumann P and Gagliardi G 2006 Dose distributions in SBRT of lung tumors: Comparison between two different treatment planning algorithms and Monte-Carlo simulation including breathing motions {\it Acta Oncol.} {\bf 45} 978-88\\[-2mm]

\noindent 
Mayorga P A, Brualla L, Sauerwein W and Lallena A M 2014 Monte Carlo study for designing a dedicated `D'-shaped collimator used in the external beam radiotherapy of retinoblastoma patients {\it Med. Phys.} {\bf 41} 011714\\[-2mm]

\noindent 
Miras H, Terr\'on J A and Lallena A M 2013 Monte Carlo simulation of COMS ophthalmic applicators loaded with Bebig I25.S16 seeds and comparison with planning system predictions {\it Phys. Med.: Eur. J. Med. Phys.} {\bf 29} 60-7\\[-2mm]

\noindent 
Ong CL, Cuijpers JP, Senan S, Slotman BJ and Verbakel WFAR 2011 Impact of the calculation resolution of AAA for small fields and RapidArc treatment plans {\it Med. Phys.} {\bf 38} 4471-9  \\[-2mm]

\noindent 
Panettieri V, Barsoum P, Westermark M, Brualla L and Lax I 2009 AAA and PBC calculation accuracy in the surface build-up region in tangential beam treatments. Phantom and breast case study with the Monte Carlo code PENELOPE {\it Radiother. Oncol.} {\bf 93} 94-101\\[-2mm]

\noindent 
Panettieri V, Wennberg B, Gagliardi G, Duch M A, Ginjaume M and Lax I 2007 SBRT of lung tumours: Monte Carlo simulation with PENELOPE of dose distributions including respiratory motion and comparison with different treatment planning systems
{\it Phys. Med. Biol.} {\bf 52} 4265-81\\[-2mm]

\noindent 
Popescu I A, Shaw C P, Zavgorodni S F and Beckham W A 2005 Absolute dose calculations for Monte Carlo simulations of radiotherapy beams {\it Phys. Med. Biol.} {\bf 50} 3375-92\\[-2mm]

\noindent 
Rodriguez M, Sempau J and Brualla L 2013 PRIMO: A graphical environment for the Monte Carlo simulation of Varian and Elekta linacs {\it Strahlenther. Onkol.} {\bf 189} 881-6\\[-2mm]

\noindent 
Rodriguez M, Sempau J and Brualla L 2015 Study of the electron transport parameters used in PENELOPE for the
Monte Carlo simulation of Linac targets {\it Med. Phys.} {\bf 42} 2877-81\\[-2mm]

\noindent 
Rogers D W O, Ma C-M, Walters B R B, Ding G X, Sheikh-Bagheri D and Zhang G 2001 BEAMnrc Users Manual PIRS-0509(A)revG (Ottawa: National Research Council of Canada); online at http://www.irs.inms.nrc.ca/inms/irs/BEAM/user manuals  2001 \\[-2mm]

\noindent 
Salvat F, Fern\'andez-Varea J M and Sempau J 2011 PENELOPE - A code system for Monte Carlo simulation of electron and photon transport (Issy-les-Moulineaux: OECD/NEA Data Bank)\\[-2mm]

\noindent 
Sauerwein W and Stannard C E 2009 Auge und orbita, in Radioonkologie Band 2: Klinik, edited by  Bamberg M, Molls M and Sack H (Munich: Zuckschwerdt Verlag), pp 294-316\\[-2mm]

\noindent 
Schipper J 1983 An accurate and simple method for megavoltage radiation therapy of retinoblastoma {\it Radiother. Oncol.} {\bf 1} 31-41\\[-2mm]

\noindent 
Schipper J, Imhoff S M and Tan K E W P 1997 Precision megavoltage external beam radiation therapy for retinoblastoma {\it Front. Radiat. Ther. Oncol.} {\bf 30} 65-80\\[-2mm]

\noindent 
Schueler A O, Fl\"uhs D, Anastassiou G, Jurklies C, Sauerwein W and Bornfeld N 2006 Beta-ray brachytherapy of retinoblastoma: feasibility of a new small-sized Ruthenium-106 plaque {\it Ophthalmic. Res.} {\bf 38} 8-12\\[-2mm]

\noindent 
Sempau J, Acosta E, Bar\'o J, Fern\'andez-Varea J M and Salvat F 1997 An algorithm for Monte Carlo simulation of coupled electron-photon transport {\it Nucl. Instrum. Meth. Phys. Res. B} {\bf 132} 377-90\\[-2mm]

\noindent 
Sempau J, Badal A and Brualla L 2011 A PENELOPE  based system for the automated Monte Carlo simulation of clinacs and voxelized geometries application to far-from-axis fields {\it Med. Phys.} {\bf 38} 5887-95\\[-2mm]

\noindent 
Sievinen J, Ulmer W and Kaissl W 2005 AAA photon dose calculation model in Eclipse$^{\rm TM}$ TRAD\#7170A (Palo Alto: Varian Medical Systems)\\[-2mm]

\noindent 
Thomson R M, Taylor R E and Rogers D W O 2008 Monte Carlo dosimetry for $^{125}$I and $^{103}$Pd eye plaque brachytherapy {\it Med. Phys.} {\bf 35} 5530-43\\[-2mm]

\noindent 
Ulmer W and Harder D 1995 A triple Gaussian pencil beam model for photon beam treatment planning {\it Z. Med. Phys.} {\bf 5} 25-30\\[-2mm]

\noindent 
Ulmer W and Kaissl W 2003 The inverse problem of a Gaussian convolution and its application to the finite size of the measurement chambers/detectors in photon and proton dosimetry {\it Phys. Med. Biol.} {\bf 48} 707-27\\[-2mm]

\noindent 
Van Esch A, Tillikainen L, Pyykkonen J, Tenhunen M, Helminen H, Siljam\"aki S, Alakuijala J, Paiusco M, Lori M and Huyskens D P 2006 Testing of the analytical anisotropic algorithm for photon dose calculation {\it Med.
Phys.} {\bf 33} 4130-48\\[-2mm]

\noindent 
Walters B R B and Rogers D W O 2003 DOSXYZnrc Users Manual PIRS-794 (Ottawa: National Research Council of Canada); 
online at http://www.irs.inms.nrc.ca/inms/irs/BEAM/user manuals\\[-2mm]

\noindent 
Zavgorodni S, Alhakeem E and Townson R 2014 Monitor backscatter factors for the Varian 21EX and TrueBeam linear accelerators: measurements and Monte Carlo modelling {\it Phys. Med. Biol.} {\bf 59} 911-24


\end{document}